\documentclass[12pt,a4paper]{article}

\usepackage[british]{babel}

\usepackage[a4paper,top=2cm,bottom=2cm,left=2.5cm,right=2.5cm,marginparwidth=1.75cm]{geometry}

\usepackage[style=apa, backend=biber]{biblatex} 
\DeclareLanguageMapping{british}{british-apa} 
\DeclareFieldFormat[article]{volume}{\apanum{#1}} 

\usepackage{amsmath}
\usepackage{graphicx}
\usepackage[colorlinks=true, allcolors=blue]{hyperref}
\usepackage{hyperref}
\usepackage[title]{appendix}
\usepackage{mathrsfs}
\usepackage{amsfonts}
\usepackage{booktabs} 
\usepackage{caption}  
\usepackage{threeparttable} 
\usepackage{algorithm}
\usepackage{algorithmicx}
\usepackage{algpseudocode}
\usepackage{listings}
\usepackage{enumitem}
\usepackage{chngcntr}
\usepackage{booktabs}
\usepackage{lipsum}
\usepackage{subcaption}
\usepackage{authblk}
\usepackage[T1]{fontenc}    
\usepackage{csquotes}       
\usepackage{diagbox}
\usepackage{float}

\usepackage{tabularx}
\usepackage[table]{xcolor}
\usepackage{pdflscape}
\usepackage{multirow}

\usepackage{setspace}
\usepackage{titlesec}
\titleformat{\section} 
  {\normalfont\Large\bfseries}{\thesection.}{1em}{}

\usepackage{lineno} 

\rightlinenumbers 

\usepackage{float}   
\usepackage{caption} 



\title{Assessing Stablecoin Credit Risks}

\author[1]{Yuval Boneh}
\author[2]{Ethan Jones}
\affil[1]{\small Conclave Head of Defi, yuvi@conclave.io}
\affil[2]{Conclave CEO, bebis@conclave.io}
\date{}  

\begin{document}
\maketitle

\begin{abstract}

  This paper delves into the spectrum of credit risks associated with decentralized stablecoin issuance, ranging from overcollateralized lending to business-to-business credit. It examines the mechanisms, risks, and mitigation strategies at each layer, highlighting the potential for scaling decentralized stablecoins while ensuring systemic health.

\end{abstract}


\pagebreak
\section{Introduction}

\begin{figure}[H]
  \centering
  \includegraphics[width=0.8\linewidth]{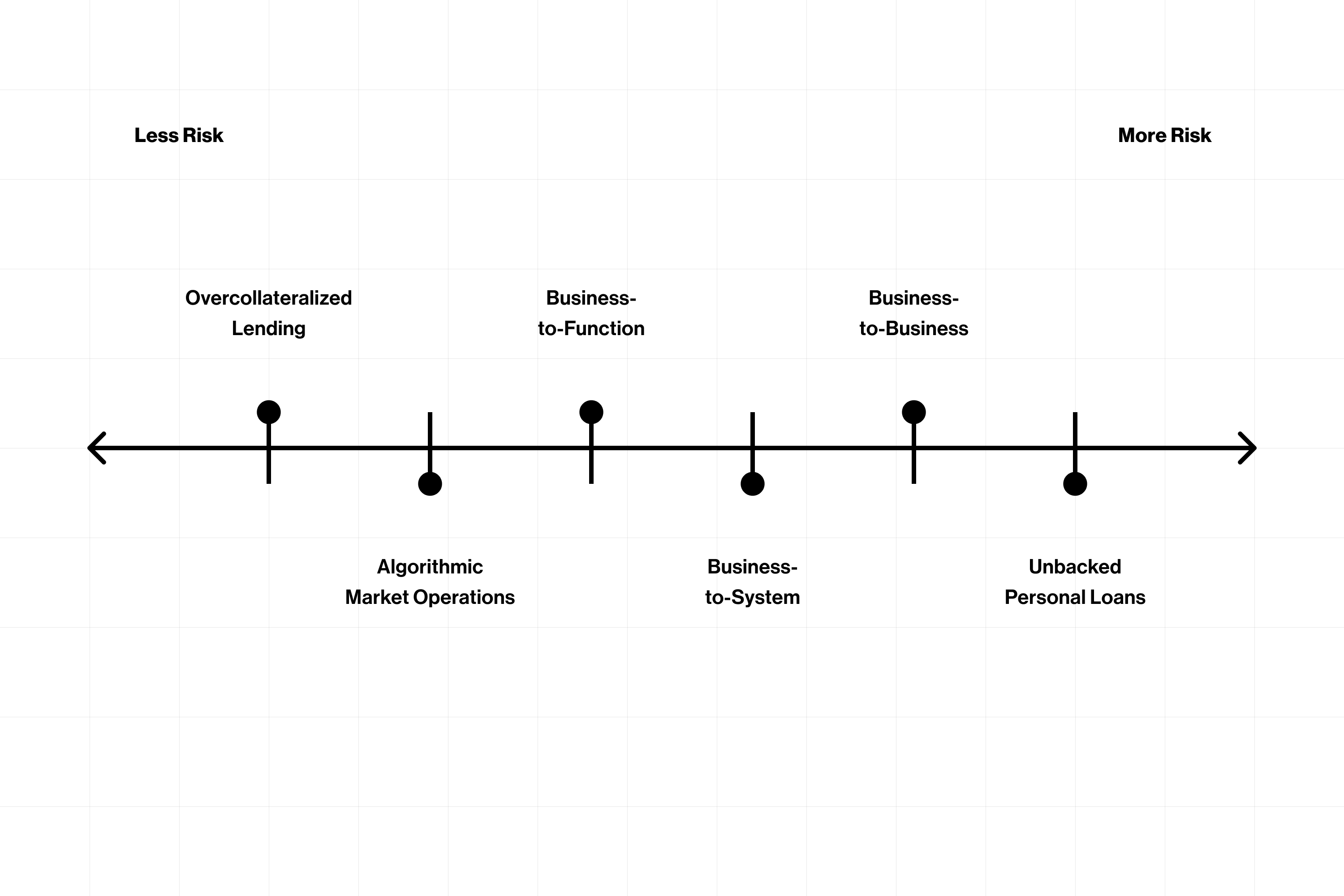}
  \caption{\label{fig:spectrum}Representation of decentralized stablecoin credit risk spectrum.}
\end{figure}

Credit risks in decentralized stablecoin issuance exist on a spectrum, with each layer introducing unique outcomes and requiring tailored mitigation strategies (Figure \ref{fig:spectrum}). Understanding these risks is crucial for the sustainable growth of decentralized stablecoins. This paper explores the spectrum from overcollateralized lending to business-to-business credit, detailing the mechanisms, risks, and mitigation strategies involved.

Risks are quantified using discretized nomenclature describing the Likelihood and Consequenece of undesirable outcomes. Each descriptor is discretized into three tiers to present a simple approach, however sophisticated risk managers and underwriters may choose to discretize further in order to better quantify the effects of mitigations. The nomenclature employed in this paper is defined in Table \ref{riskQuant}.

\begin{table}[H]
\caption{Risk level definitions\label{riskQuant}}
\begin{threeparttable}
\begin{tabularx}{\columnwidth}{@{}p{2.1cm}p{2.2cm}X@{}}
\toprule
Descriptor & Level & Definition\\
\midrule
\multirow{3}{*}{Likelihood} & Low (A) & Not expected to occur.\\
& Medium (B) & May foreseeably occur.\\
& High (C) & Certain to occur.\\
\\
\multirow{3}{*}{Consequence} & Low (1) & Marginal, temporary impact to systemic health.\\
& Medium (2) & Considerable, temporary impact to systemic health.\\
& High (3) & Existential impact to systemic health.\\
\bottomrule
\end{tabularx}
\end{threeparttable}
\end{table}

A summary of the risks identified in this paper can be located in Appendix \ref{appendix:risk_matrix}, and should be viewed in conjunction with the paper itself.

\subsection{Scope}

For most types of issuance expressed in this paper there exists a certain level of trust in the mechanisms that govern risk and return when a new line of credit is underwritten. In the simplest example, Uniswap V2, the AMM function of the platform will always abide by the invariant x*y=k. On Aave, supply and demand is managed by Loan-To-Value (LTV) ratios and interest rate mechanisms.

In underwriting lines of credit to functions or systems, security and risk transfer must be handled entirely by these algorithms. In certain cases, extreme capital efficiency can be achieved by using these algorithms to define and govern risk relationships between counterparties.

This paper explores risks introduced specifically by the credit functions described and does not delve into extant risks beyond the scope of stablecoin credit issuance. Certain dependency risks, such as security exploits, are not unique to credit issuance and are considered out-of-scope for the purpose of this paper. Despite this, a thorough understanding of each mechanism, their potential edge-cases, and the means through which they can be modified or invalidated is required to properly underwrite defi loans.

\section{Overcollateralized Lending}

\begin{figure}[H]
  \centering
  \includegraphics[width=0.8\linewidth]{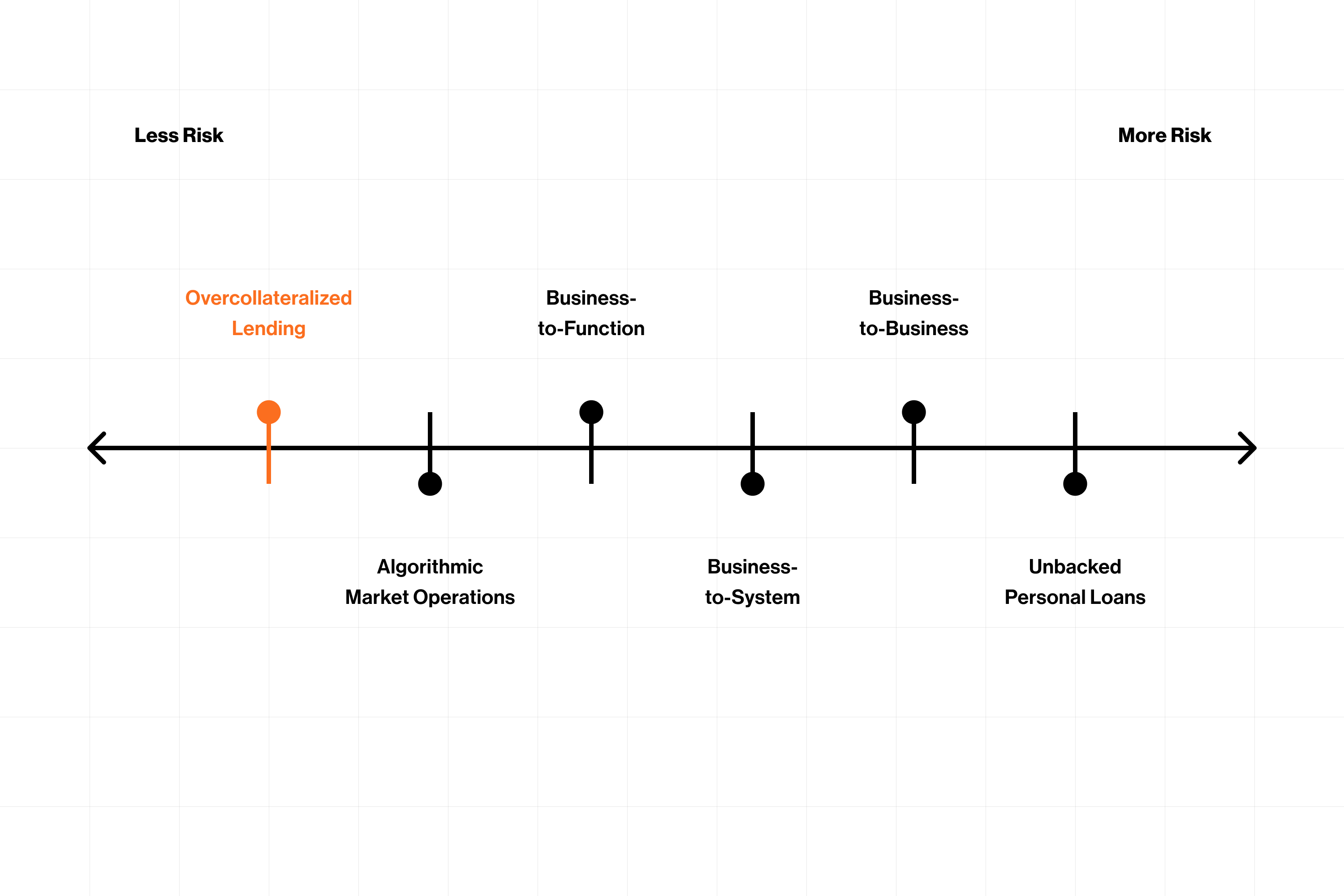}
  \caption{\label{fig:spectrumOver}Overcollateralized Lending on the decentralized stablecoin credit risk spectrum.}
\end{figure}

The lowest risk level of credit issuance is overcollateralized lending (Figure \ref{fig:spectrumOver}), as seen in protocols that rely on collateralized debt positions. Stablecoins cannot be minted, in or out of circulation, unless they are collateralized by assets, autonomously underwriting the loan. The primary risk is that the value of the asset backing the loan depreciates to below the loan value, referred to as Liquidation Risk.

\subsection{Aave and GHO}

Aave's GHO stablecoin uses overcollateralization principles to maintain a stable value. "GHO is minted upon the supply of crypto-assets in excess of the value of GHO to be minted" (\cite{AaveGho}). The value of debt compared to collateral is referred to as the Health Factor, and accounts for the Liquidation Threshold. If a user deposits \$10,000 worth of a collateral asset that has a Liquidation Threshold of 80\%, the maximum amount of \$GHO they would be able to borrow (ignoring Loan-To-Value Ratio limitations) is \$8,000. At this point, their Health Factor would be 1, and if the value of their collateral dropped below \$10,000, they would be at risk of liquidation. If liquidated, their collateral assets would be sold to recover the debt and part or all of the difference in value would be retained by the liquidator. This incentivizes liquidators to keep the protocol healthy, and disincentivizes borrowers from excessively leveraging their collateral.

Overcollateralized lending typically transfers Liqudiation Risk to asset suppliers, who are compensated by receiving a majority of the interest paid by asset borrowers. In the case of \$GHO, Aave maintains risk ownership by minting \$GHO to the borrower instead of having users supply \$GHO. In turn, Aave keeps the interest paid on \$GHO borrowed.

The Aave DAO manually adjusts debt limits and interest rates to ensure the value of \$GHO remains pegged to \$1. If users borrow \$GHO and sell it excessively and the price drops, the Aave DAO will vote to increase the interest rate, pressuring borrowers to buy back \$GHO to repay their loans. The same occurs in the other direction, whereby the Aave DAO may vote to decrease rates in order to expand the supply of \$GHO.

\subsection{Liquidation Risk}

Liquidation Risk is perceived to have High likelihood since leverage is the core purpose of such a product, and Medium consequence, since it is not necessarily an existential threat. This risk has been realized historically by CDPs and lending protocols and losses are often covered by the protocol. This risk is mitigated upfront with adequate interest rates, and further mitigated with robust liquidation and redistribution infrastructure, ensuring a margin of safety defined by the Loan-To-Value ratio, liquidation threshold, liquidation bonus, etc.

These mitigations reduce the Likelihood of the risk eventuating by compensating issuers for assuming risk (where applicable) and liquidators for liquidating unhealthy positions before bad debt is accrued and unbacked stablecoins can circulate.

\section{Algorithmic Market Operations}

\begin{figure}[H]
  \centering
  \includegraphics[width=0.8\linewidth]{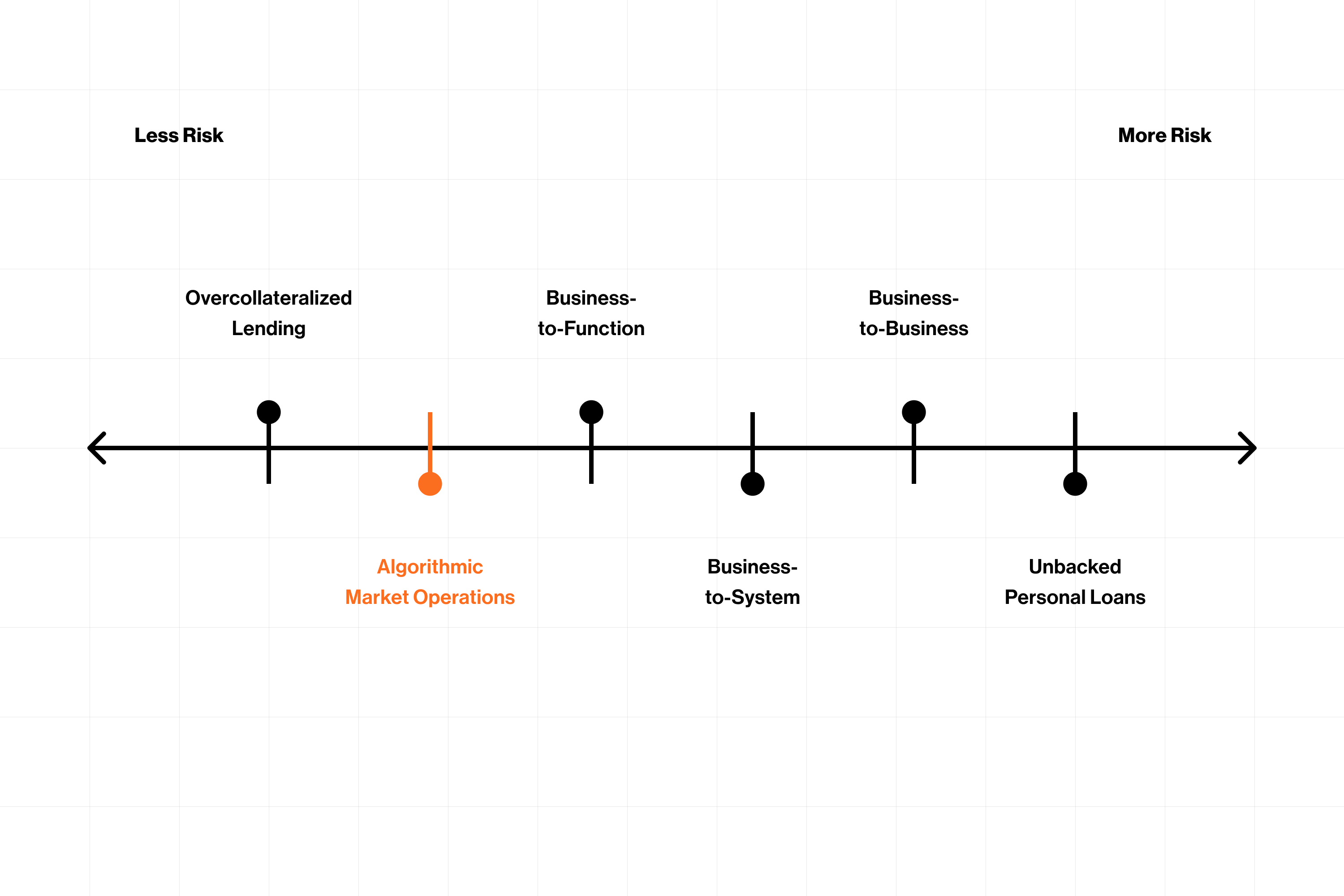}
  \caption{\label{fig:spectrumAmos}Algorithmic Market Operations on the decentralized stablecoin credit risk spectrum.}
\end{figure}

Algorithmic Market Operations (AMOs) are autonomous contracts that enact arbitrary monetary policy according to their mathematical programming. Typically speaking, some amount of stablecoins is minted and custodied by an autonomous system such that the liquidity exists but is functionally out of circulation. That is, the AMO with custody of liquidity introduces more risk (Figure \ref{fig:spectrumAmos}) but cannot have a net effect on liquidity conditions to the detriment of systemic health.

\subsection{Peg Stability Module}

\begin{figure}[H]
  \centering
  \includegraphics[width=0.8\linewidth]{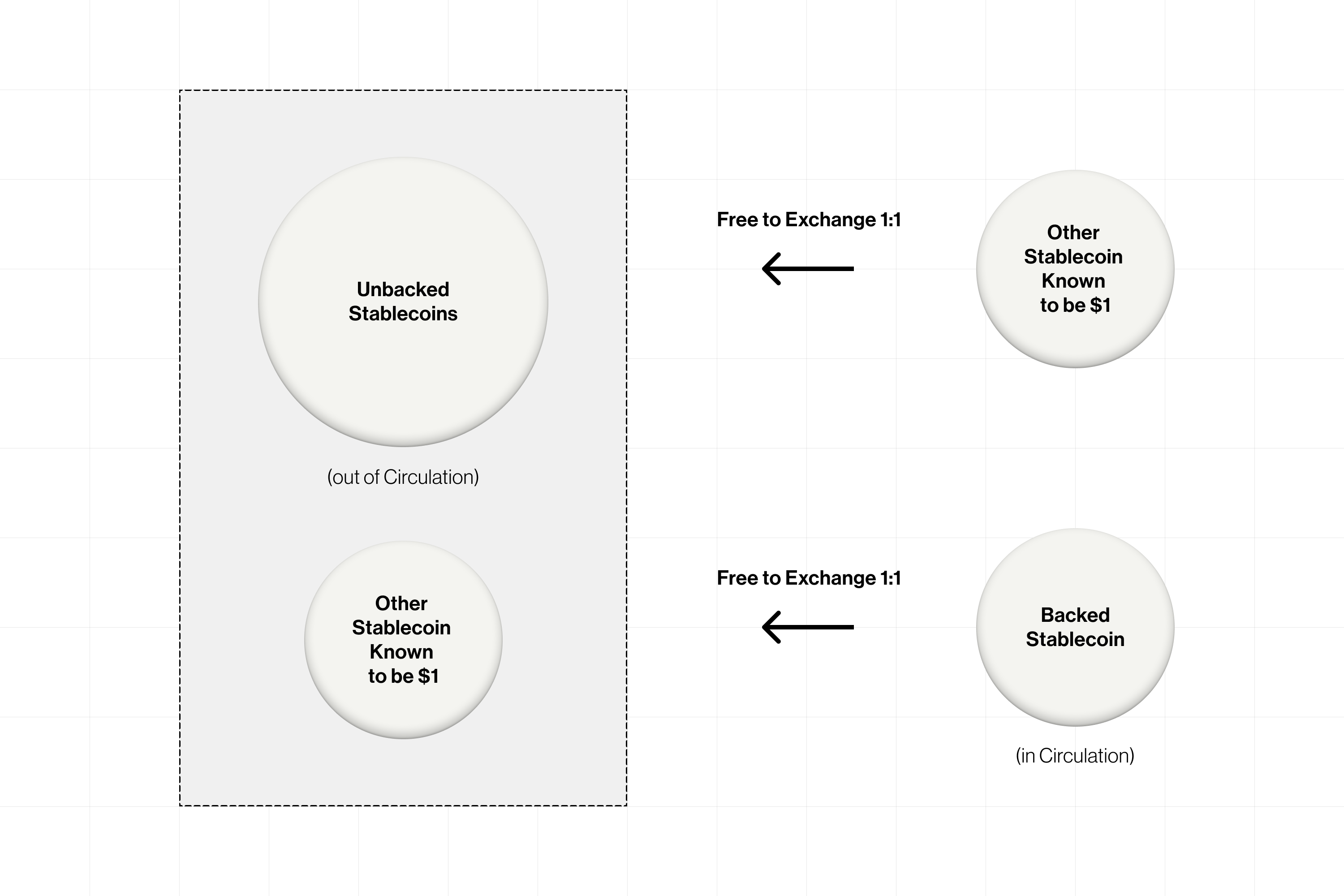}
  \caption{\label{fig:psm}Representation of a simple Peg Stability Module.}
\end{figure}

In the case of a Peg Stability Module (PSM), the system serves as a large reserve of unbacked stablecoins that facilitates arbitrage operations to improve peg stability (Figure \ref{fig:psm}). Supply is pulled and pushed by external users, and the arbitrage counterassets effectively back any liquidity that subsequently enters circulation.

\subsection{Liquidity AMO}

\begin{figure}[H]
  \centering
  \includegraphics[width=0.8\linewidth]{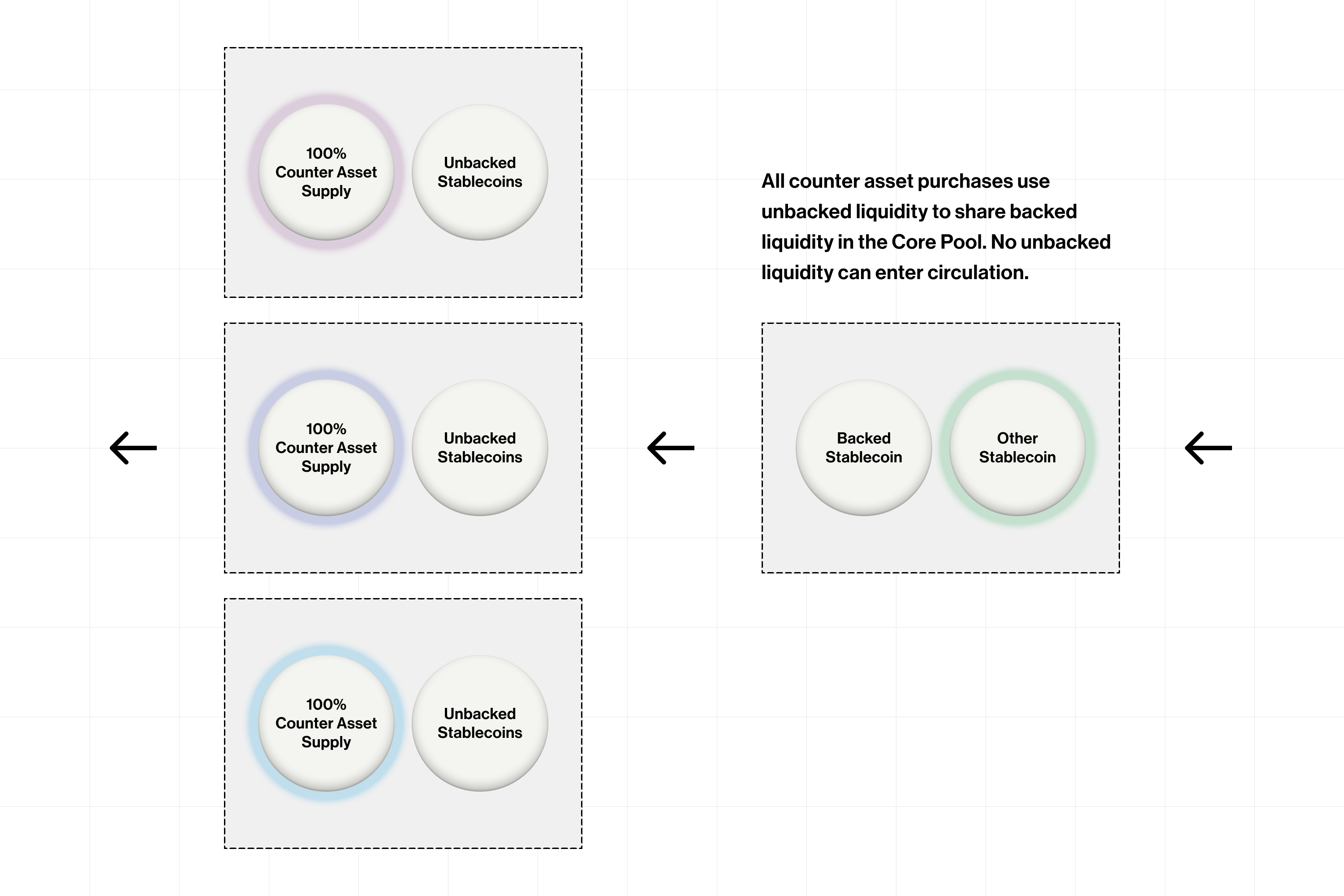}
  \caption{\label{fig:liquidityAmo}Representation of a Liquidity AMO.}
\end{figure}

For Liquidity AMOs, 100\% of the counterasset supply is pooled with unbacked stablecoins (\cite{cdxUSD}). Since there are no circulating counterassets outside of the pool, there is no way to sell into the stablecoin and draw it into circulation (Figure \ref{fig:liquidityAmo}). Therefore, the only way to access the stablecoin is from another facilitator, whether it is backed by a loan or purchased from another liquidity pool and stabilized by a PSM. The stablecoin cannot enter circulation unless it is effectively backed.

\subsection{Operations Risk}

The key risk introduced by this layer of functionality is Operations Risk. AMOs rely on code correctness and external dependencies in order to safely execute their functions. if misconfigured or mismanaged, Mint and burn functions, transfer functions, and other administrative or guardian functions may execute entirely as programmed while resulting in unbacked stablecoins entering circulation 

Whether by accident or through nefarious social or engineering outcomes, the likelihood of this risk is considered to be Medium, with existential consequences (High).

This risk is often mitigated in the interim with multisigs and timelocks, and more enduringly by programming safeguards and iterating towards permissionless and immutable functionality wherever possible. These mitigations reduce the likelihood to Low and in some cases may reduce the consequence to Medium.

\section{Business-To-Function (B2F) Credit}

\begin{figure}[H]
  \centering
  \includegraphics[width=0.8\linewidth]{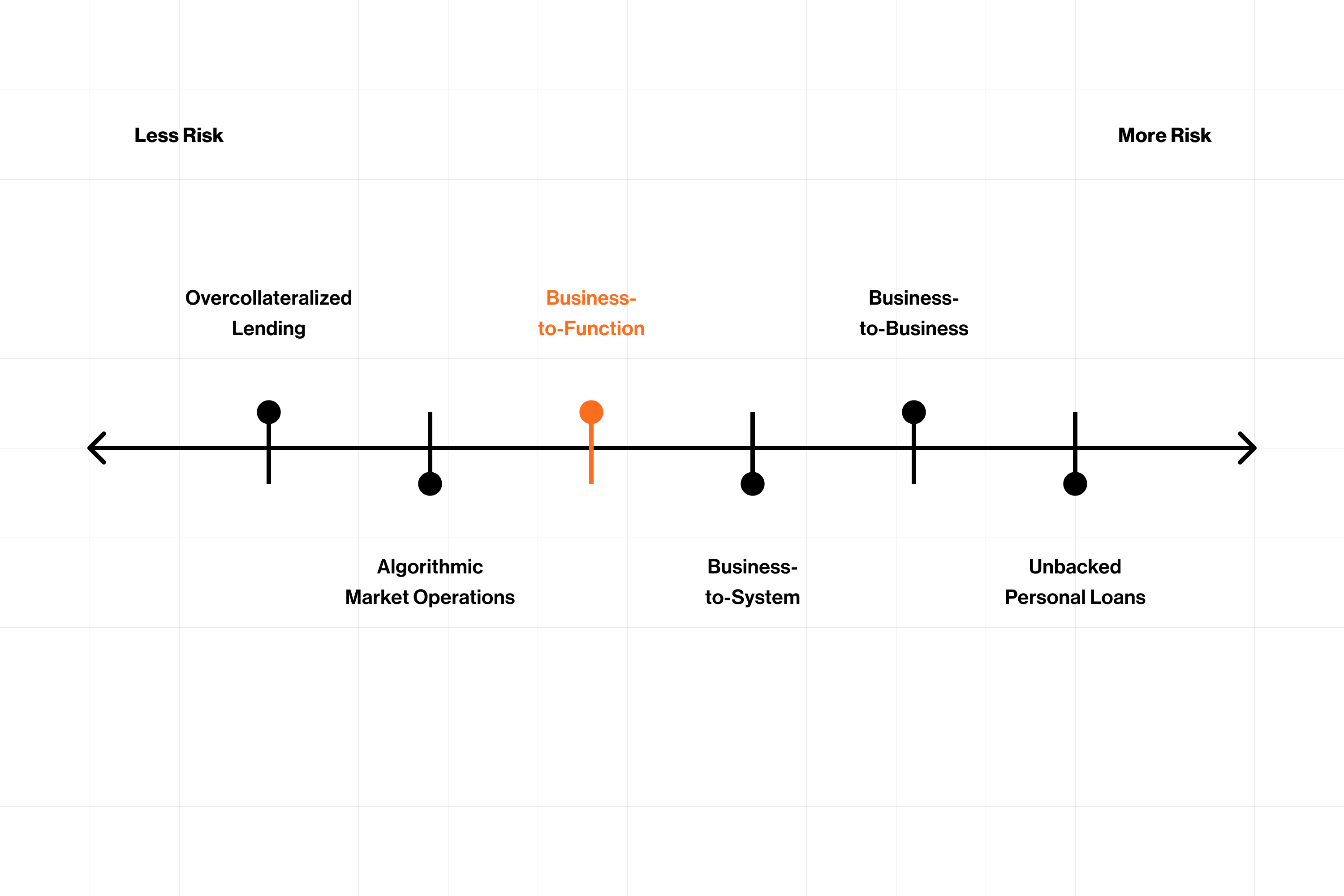}
  \caption{\label{fig:spectrumB2f}B2F on the decentralized stablecoin credit risk spectrum.}
\end{figure}

Further along the risk spectrum, unbacked stablecoins may be issued to smart contract processes that allow assets into circulation, where there is a cost of liquidity (Figure \ref{fig:spectrumB2f}). For example, an unbacked sum of stablecoins may be minted and issued to an external isolated lending market. While the assets supplied are initially not circulating, they may enter circulation if a user deposits collateral to borrow the stablecoin. The implications here are that the external lending market's liquidation infrastructure is inherited and must be able to handle the volume of collateral accepted and loans issued, and that the cost of borrowing (interest rate) does not negatively impact the stablecoin's systemic health.

Liquidation Risk is inherited from the external protocol's liquidation infrastructure and requires independent assessment. The consequences of bad debt may be further mitigated by limiting the amount of debt issued to the external lending market.

\subsection{Maker's DAI/sUSDe Morpho Vault}

The largest example of stablecoin credit issuance is Maker's DAI Direct Deposit Module (DDM, \cite{DDM}), a tool that directly injects DAI (Maker's stablecoin) into third party protocols.

In March 2024, MakerDAO identified a hedged perpetual yield opportunity allowing them exposure to Ethena's yield-bearing sUSDe product without depending on a custodian (\cite{SkyGov}). The proposal involved injecting DAI liquidity into a Morpho vault with USDe and sUSDe counterassets. External sUSDe holders could then deposit sUSDe and borrow DAI for purchasing more sUSDe. The intended effect is that external holders increase their sUSDe exposure as DAI utilization increases, while Maker benfits from the interest paid on DAI.

BA Labs conducted a Risk Assessment for the USDe Morpho Lending Integration (\cite{BALabs}) and aptly identified a suite of risks ranging from Exchange Failure and Funding Rates (specific to the Ethena integration), to lending efficiency and debt cap parameters (specific to Maker's DAI injection).

This particular example forms an excellent basis to consider the risks associated with uncollateralized lines of credit.

\subsection{Cost of Borrowing Risk}

Cost of borrowing introduces a new risk. In cases where on-chain liquidity affects other facilitators, the line of credit to any such market should be limited based on liquidity conditions and interest rate characteristics. The effects of this risk can be mitigated by understanding the interest rate model and curating a market that reflects the expected asset demand. The premise of the DAI/sUSDe Vault is that DAI would be borrowed and sold in favour of sUSDe, such that the cost of borrowing DAI would reach parity with sUSDe yield. This is determined by the size of credit issued and the interest rate characteristics implemented.

In order to underwrite a B2F line of credit, liquidity conditions related to the Facilitator should satisfy Equation \ref{irmComp}:

\begin{equation}
  cost_{external}(x) \geq yield_{facilitator}(x),
  \label{irmComp}
\end{equation}

where $x$ represents some volume of liquidity. In context, the cost of borrowing $x$ from the external protocol should be greater than the yield of depositing $x$ into the facilitator. Otherwise, users of the facilitator face an opportunity cost deficit to those extracting from external credit lines.

Since liquidity conditions and market dynamics are transient, the likelihood of encountering an undesirable event is High, however yield parity and rate arbitrage will foreseeably limit consequences to Medium. Active monitoring infrastructure that can respond to changes in liquidity conditions within reasonable operating bounds may reduce both Likelihood and Consequence.

By understanding the dynamics of facilitator yield in relation to liquidity conditions, an appropriate size and cost of credit can be established to suit the operating parameters of the external protocol.

\subsection{Cost of Borrowing Risk: Example}\label{example}

The Cod3x stablecoin $cdxUSD$ (\cite{cdxUSD}) utilizes a core liquidity pool to ascertain market demand block-to-block, to control the interest rate that borrowers pay (\cite{MoneySup}) through its primary issuance Facilitator, Cod3x Lend (\cite{cdxLend}).

In the example described by \textcite{MoneySup}, a Curve StableSwap pool with amplifiication factor set to 100, and with 1,000,000 cdxUSD and 1,000,000 stablecoin counterassets, can facilitate a swap of 400,000 cdxUSD for approximately 398,132 counterassets, resulting in a pool balance of
approximately 70\% cdxUSD, at a price of approximately \$0.995.

Following the example architecture, the controller is tuned to output an interest rate of 10\%, thus determining the cost of liquidity (facilitator yield) in conditions of lowest demand.

In order to satisfy Equation \ref{irmComp}, the cost of borrowing 400,000 cdxUSD from an external facilitator should be greater than 10\%. This way, if a user borrows 400,000 cdxUSD to sell into the core liquidity pool, they will pay a higher interest rate than Lend users, who could then arbitrage the yield or repay their debt at a discount (ingoring extraneous arbitrageurs).

An Aave market Facilitator with optimal utilzation set to 80\% and VariableRateSlope1 set to 1e26 would result in an interest rate of 10\% at the optimal utilization target (\cite{AaveIRM}). Therefore, a line of credit of up to 500,000 cdxUSD could reasonably be afforded to an Aave market Facilitator without negatively impacting Cod3x Lend.

Equation \ref{irmComp} is contextualized by comparing the interest rates ($IR$) of the underlying facilitators, per Equation \ref{irmComp2}.

\begin{equation}
  IR_{Aave} \geq IR_{Lend},
  \label{irmComp2}
\end{equation}

The transfer function derived by \textcite{MoneySup} is provided in Equation \ref{irmComp3}.

\begin{equation}
  r = 0.15 * \frac{E_{controller}}{1 - E_{controller}}\\
  \label{irmComp3}
\end{equation}

In order to limit the line of credit safely, it is assumed that Aave utilization is sold to the market, represented in Equation \ref{irmComp4}.

\begin{equation}
  E_{controller} = X * U\\
  \label{irmComp4}
\end{equation}

where $X$ is the size of credit offered relative to the counterassets in the core liquidity pool. This is because if some cdxUSD has already been sold, there will be a higher quantity of cdxUSD in the pool, and so a smaller line of credit should be afforded, effectively scaling down with the smaller number of counterassets in the core pool.

Standard Aave interest rate curves can then be compared to the transfer function in Equation \ref{irmComp5}. In this case, the equation is defined at the instance of utilization and does not account for the adaptive nature of the Cod3x Lend controller, since market conditions over time are inherently invalidated. Additionally, the Aave interest rate manager base rate is set to a conservative zero.

\begin{equation}
  \begin{cases}
  \frac{U}{U_{optimal}}*R_{slope1}, & U \in (0, U_{optimal}] \\
  R_{slope1} + \frac{U-U_{optimal}}{1 - U_{optimal}}*R_{slope2}, & U \in (U_{optimal}, 1)
  \end{cases}
  \geq 0.15 * \frac{X * U}{1 - X * U}
  \label{irmComp5}
\end{equation}

If utilization is maintained at optimal, and setting an arbitrary optimal utilization of $0.8$, then the comparison simplifies to Equation \ref{irmComp6}.

\begin{equation}
  R_{slope1} \geq \frac{3*X}{25 - 20*X}
  \label{irmComp6}
\end{equation}

If the desired rate at optimal utilization is 10\%, then $R_{slope1} = 0.1$, and $X$ resolves to 0.5. This means that for a \$2,000,000 pool with 1,000,000 cdxUSD and 1,000,000 USDC, a 500,000 cdxUSD line of credit can reasonably be afforded to an Aave market.

Since the Cod3x Lend cdxUSD interest rate is adaptive and responds to market dynamics, the line of credit afforded to an external Facilitator should be regularly revised to account for shifts in market demand. This example provides a baseline assessment for underwriting credit, however further consideration regarding the performance of Arbitrage AMOs and other sources of demand may justify increasing the line of credit.

\subsubsection{Subsequent Mitigation: Endogenous Yield}\label{endoYield}

Assuming conditions are satisifed such that the line of credit issued to the lending market is generating yield (borrowers paying interest, less the protocol reserve factor), the yield generated can then be directed to support $cdxUSD$ liquidity. This is a second order effect that serves to further mitigate Cost of Borrowing Risk.

At optimal utilization, \$400,000 paying 10\%, and assuming a reserve factor of 20\%, means \$32,000 of interest paid in one year. Similarly, 10\% paid on \$1,000,000 of cdxUSD issued via Cod3x Lend, that is providing liqudity to the core pool, means \$100,000 of interest paid yearly. Directing \$132,000 to the \$2,000,000 liquidity pool sustains 6.6\% yield, not accounting for trading fees.

This offsets the cost of borrowing from Cod3x Lend and inherently mitigates the Cost of Borrowing Risk by reducing the likelihood of mismatched costs. The more $cdxUSD$ that is sold, the higher the interest that is paid, and yield that is directed to support liquidity endogenously and sustainably.

\section{Business-To-System (B2S) Credit}

\begin{figure}[H]
  \centering
  \includegraphics[width=0.8\linewidth]{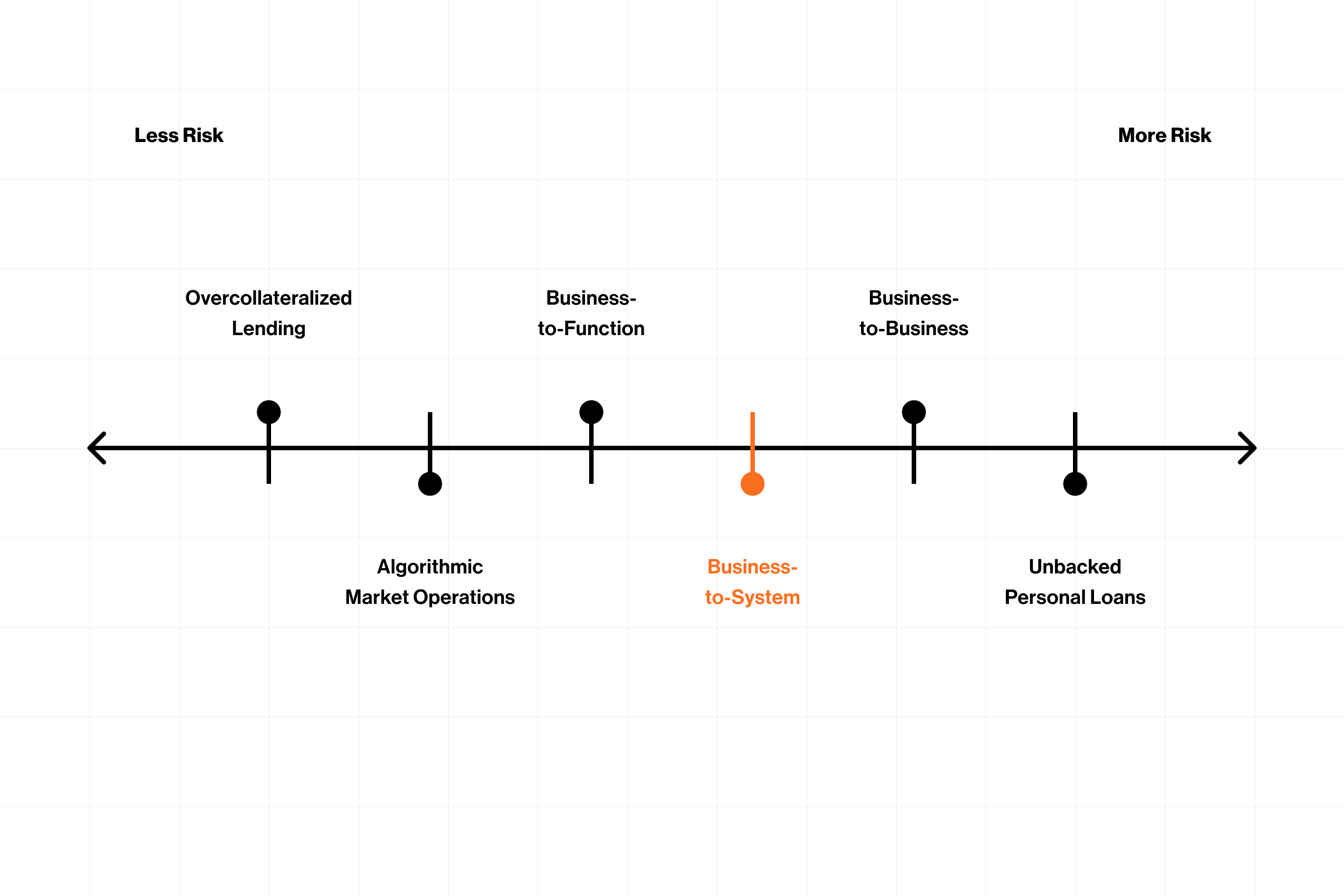}
  \caption{\label{fig:spectrumB2s}B2S on the decentralized stablecoin credit risk spectrum.}
\end{figure}

The key differentiator between a B2F loan and a B2S loan is that the credit issuer has no control of the cost for accessing unbacked stablecoins. Issuing credit to a system implies a layer of complexity beyond that of an isolated function (Figure \ref{fig:spectrumB2s}) such that the context and extraneous variable have significant influence on the subsequent circulating supply (typically at no ongoing cost).

\subsection{Perpetuals Exchanges}

Allowing unbacked assets into circulation at no cost requires more nuanced justification. An ideal example of a B2S loan would be the provision of counterparty liquidity to a perpetuals exchange. In this case, there is no cost to utilize the liquidity (less funding rates), however the issuer's incentive is yield derived from traders.

Decentralized perpetuals exchanges often develop vaults to attract counterparty liquidity for traders. This typically means that when traders win or lose, the liquidity that balances the books is drawn from the counterparty pool. The attractiveness comes from the statistical probability that over a long enough time horizon, traders tend to lose money, thus growing the counterparty pool and providing counterparties with a source of yield.

The size of the counterparty pool often plays a role in the exchange's open interest and funding limitations, so exchanges and traders alike want deep liquidity. However, asset supply is typically limited by risk-adjusted interest rate parity and the opportunity cost of depositors. To overcome this, it is possible to issue an unbacked line of credit to a perpetuals exchange to facilitate a large deposit of counterparty stablecoins.

\subsection{Unbacked Circulation Risk}

The risks to this point are largely unchanged since the vault is programmatically controlled. However, this functionality has the potential to introduce unbacked stablecoins into circulation with no cost to the holder. That is, for any period that traders are winning, they may take possession of circulating, unbacked stablecoins.

The difference in this Unbacked Circulation Risk is that since there is no cost of accessing liquidity, the only controllable mitigation is to limit the credit issued to the counterparty vault. If the exchange is established, historical data may be assessed to ascertain the duration and volumes that traders have typically won and realized profits. By understanding this data, it is reasonable to increase the loan size and allow the entropy of trading to mitigate these risks. In that case, a stablecoin counterasset fund should be established to temporarily backfill any realized profits that significantly affect systemic health via the core liquidity pool. This may also be achieved with arbitrage.

Structurally mitigating Unbacked Circulation Risk is impossible because neither stakeholder can influence the performance of traders. The circulation of unbacked stablecoins at some point in time is practically guaranteed, however the worst case scenario of the quantity of unbacked stablecoins being sufficient to cause High level consequences is far less likely (Low) if credit is appropriately sized.

\subsection{Unbacked Circulation Risk: Example}

The decentralized perpetuals exchange Gains Network utilizes collateral vaults as counterparties to all trades made on the platform (\cite{Gains}). Historically, vault collateralization has dropped below 100\% rarely, generally only very early on in the vault's lifecycle, and by no more than 6\% (\cite{GainsStats}), meaning that practically speaking, unbacked circulation by way of traders winning on perpetuals exchanges is extremely rare.

Another consideration is that a $cdxUSD$ counterparty vault would require traders to collateralize their trades with $cdxUSD$, meaning that traders acquire $cdxUSD$ with which to trade. Traders who buy cdxUSD from the market effectively provide liquidity by way of cdxUSD pool counterassets, deepening liquidity and mitigating the consequences of unbacked circulation risk. Traders who borrow cdxUSD with which to trade are therefore incentivized to repay their loans with their winnings, as opposed to selling winnings to other assets, reducing the likelihood of unbacked circulation. In some capacity, traders who utilize cdxUSD on the perpetuals exchange are providing an interest-free loan to the cdxUSD infrastructure while they wait for their trades to play out.

The two key guiding indicators for sizing a line of credit are the appetite for traders to utilize cdxUSD as their denominated trading asset, and the rate at which traders are winning or losing.

Gains Network's gUSDC vaults are earning an average of 11.75\% (\cite{GainsStats}), which fluctuates over time, and can be treated as 10\% for the purpose of this example. Following the logic in \ref{example}, the core liquidity pool can absorb \$400,000 of cdxUSD before Lend rates rise to 10\%. Assuming an equal worst case of 6\% undercollateralization, a \$6.67M line of credit could facilitate the expected trading behaviour, noting that the risk reduces over time, and not accounting for volume of cdxUSD purchased by traders.

This also does not account for the subsequent order effect of endogenous yield supporting liquidity as highlighted previously in \ref{endoYield}.

This is typically the extent to which decentralized stablecoin issuers underwrite lines of credit because the risks are understood and the liability transferred to smart contract processes.

\section{Business-to-Business (B2B) Credit}

Beyond programmatic custody on the risk spectrum exists business-to-business credit, whereby a decentralized stablecoin issuer facilitates an unbacked loan of stablecoins to a market maker or trading firm to generate yield. At this end of the spectrum, the risks are contextual and in some cases extraneous.

Understanding the myriad of risks introduced by this approach is out of scope of this research. However, ensuring that stakeholders document and monitor risks and interim and enduring mitigations should remain a core practice.

B2B credit should typically be underwritten contractually and with appropriate legal enforcement actions understood by both parties, however the on-chain implications are not clearly understood.

\section{Conclusion}

The spectrum of credit risks in decentralized stablecoin issuance ranges from overcollateralized lending to business-to-business credit, and beyond. Each layer typicaly inherets the risk profile of the layer beneath it, and introduces unique risks that require additional mitigation strategies. By understanding and addressing these risks, decentralized stablecoins can scale sustainably, opening up tremendous opportunities for growth and innovation. The potential for scaling decentralized stablecoins by issuing lines of credit to business processes, when risks can be assessed and sufficiently mitigated, is substantial. This paper has explored the mechanisms, risks, and mitigation strategies at each layer, highlighting the importance of understanding and managing credit risks for the sustainable growth of decentralized stablecoins.

\printbibliography

\pagebreak

\appendix

\renewcommand\theequation{\Alph{section}\arabic{equation}} 
\counterwithin*{equation}{section} 
\renewcommand\thefigure{\Alph{section}\arabic{figure}} 
\counterwithin*{figure}{section} 
\renewcommand\thetable{\Alph{section}\arabic{table}} 
\counterwithin*{table}{section} 

\begin{appendices}
\begin{landscape}
\section*{Appendix}
\section{Risk Matrix}
\label{appendix:risk_matrix}

Table \ref{tab_a1} summarizes stablecoin credit risks.

\begin{table}[H]
\caption{Decentralized stablecoin credit issuance risk matrix\label{tab_a1}}
\begin{threeparttable}
\begin{tabularx}{\columnwidth}{@{}p{4.8cm}p{2.2cm}p{2.1cm}XXp{2.2cm}p{2.1cm}@{}}
\toprule

& \multicolumn{2}{c}{Unmitigated} & & & \multicolumn{2}{c}{Mitigated}\\
\cmidrule(lr){2-3}\cmidrule(lr){6-7}
Risk & Likelihood & Consequence & Interim Mitigations & Enduring Mitigations & Likelihood & Consequence\\
\midrule
Liquidation Risk & High (C) & Medium (2) & System Health Monitoring & Liquidation/ Redistribution Infrastructure & Low (A) & Medium (2)\\
Operations Risk & Medium (B) & High (3) & Multisignature Safe, Timelock & Immutable, Permissionless & Low (A) & Medium (2)\\
Cost of Borrowing Risk & High (C) & Medium (2) & Excessive Rates, Credit Limit & Active Monitoring Infrastructure, Endogenous Yield & Medium (B) & Medium (2)\\
Unbacked Circulation Risk & Medium (B) & High (3) & Credit Limit & Active Monitoring Infrastructure, Endogenous Yield & Low (A) & High (3)\\
\bottomrule
\end{tabularx}
\end{threeparttable}
\end{table}

\newpage

Table \ref{tab_a2} visually presents quantified risk levels.

\begin{table}[H]
\caption{Risk quantifying matrix\label{tab_a2}}
\begin{threeparttable}
\begin{tabularx}{\columnwidth}{@{}X|XXXX@{}}
\toprule
Descriptor & \multicolumn{4}{c}{Consequence}\\
\midrule
\multirow{4}{*}{Likelihood} & & Low (1) & Medium (2) & High (3)\\
& Low (A) & \cellcolor[HTML]{0000FF}A1 & \cellcolor[HTML]{00ff00}A2 & \cellcolor[HTML]{FFFF00}A3\\
& Medium (B) & \cellcolor[HTML]{00ff00}B1 & \cellcolor[HTML]{FFFF00}B2 & \cellcolor[HTML]{FFA500}B3\\
& High (C) & \cellcolor[HTML]{FFFF00}C1 & \cellcolor[HTML]{FFA500}C2 & \cellcolor[HTML]{FF0000}C3\\
\bottomrule
\end{tabularx}
\end{threeparttable}
\end{table}

\end{landscape}
\end{appendices}

\end{document}